\documentclass[prc,showpacs,twocolumn]{revtex4}
\usepackage{graphicx}
\usepackage{dcolumn}
\usepackage{bm}

\begin{document}

\title{Microscopic description of fission in  
nobelium isotopes  
with the Gogny-D1M 
energy density functional}

\author{R. Rodr\'{\i}guez-Guzm\'an}
\affiliation{Physics Department, Kuwait University, Kuwait 13060, Kuwait}

\author{L.M. Robledo}
\affiliation{Departamento  de F\'{\i}sica Te\'orica, 
Universidad Aut\'onoma de Madrid, 28049-Madrid, Spain}

\date{\today}

\begin{abstract}
Constrained mean-field calculations, based on the Gogny-D1M energy 
density functional, have been carried out to describe fission in the 
isotopes $^{250-260}$No. The even-even isotopes have been considered 
within the standard Hartree-Fock-Bogoliobov (HFB) framework while for 
the odd-mass ones  the Equal Filling Approximation (HFB-EFA) has been 
employed. Ground state quantum numbers and deformations, pairing 
energies, one-neutron separation energies, inner and outer barrier 
heights as well as fission isomer excitation energies are given. 
Fission paths, collective masses and zero-point quantum vibrational and 
rotational  corrections are used to compute the systematic of the 
spontaneous fission half-lives t$_\mathrm{SF}$ both for even-even and 
odd-mass nuclei. Though there exists a strong variance of the predicted 
fission rates with respect to the details involved in their 
computation, it is shown that both the specialization energy  and the 
pairing quenching effects, taken into account within the self-consistent 
HFB-EFA blocking procedure, lead to larger  t$_\mathrm{SF}$ values in 
odd-mass nuclei as compared with their even-even neighbors. Alpha decay 
lifetimes have also been computed using a  parametrization of the 
 Viola-Seaborg formula. 
The high quality of the Gogny-D1M functional regarding nuclear masses leads to a very good 
reproduction of $Q_{\alpha}$ values and consequently of lifetimes.
\end{abstract}


\maketitle{}

%
%
%

\section{Introduction}

The theoretical description of odd-mass nuclei  still remains as a 
major challenge in nuclear structure as compared to the treatment of 
even-even nuclei. Nevertheless, odd-mass nuclei provide key information  regarding 
not only odd-even effects associated to pairing correlations but also the 
``time-odd" section of the nuclear energy density functional (EDF) which is of 
paramount relevance in the dynamical behavior  of the nucleus. In addition, 
odd-mass nuclei could also be useful  to constrain the parameters of 
energy density functionals (EDFs)  with the aim to reach a reasonable spectroscopic 
quality. Moreover, the energies, spins and parities of their low-lying 
one-quasiparticle spectra provide  insight into the underlying shell 
effects. Thus, the properties of odd-mass nuclei and the 
computational challenge in their evaluation have received renewed attention in recent 
years (see, for example, 
\cite{Hamamoto-dd,EFA-jsut,EFA-Bonneau,duguet-odd,EFA-Rayner-1,EFA-Rayner-2,EFA-Rayner-3,EFA-Rayner-4,reorient-1,reorient-2,No-REGION} 
and references therein).

In the context of spontaneous fission \cite{Bjor,Specht,Holden-paper}, 
it is experimentally observed that the  half-lives t$_\mathrm{SF}$ of 
odd-mass nuclei are systematically  larger than the ones of their 
even-even neighbors. Two complementary mechanisms have been proposed to 
explain this feature: One is the specialization energy 
\cite{special-Fong}, that modifies the collective potential energy 
landscape felt by the odd-mass nucleus in its evolution from the ground 
state to scission as compared to the one of the neighboring even-even 
companions. The origin of part of this energy is the assumption that 
the K quantum number associated with the projection of the angular 
momentum onto the intrinsic symmetry axis characteristic of any 
odd-mass quantum state has to be conserved along the fission process. 
The reason behind this assumption is that the time scale for fission 
(not to be confused with the t$_\mathrm{SF}$ value) is of the order of 
the one for the strong interaction and, therefore, much shorter than 
the typical time scales for the electromagnetic processes responsible 
for the change of the K  quantum number. The second mechanism  is 
the characteristic quenching of  pairing correlations caused by the 
unpaired nucleon that weakens the pairing field strengths, as compared 
with the fully paired situation, enhancing the collective inertia that 
strongly depends on the pairing gap. This enhancement in turn, 
increases the collective action and, therefore, the spontaneous fission 
half-life t$_\mathrm{SF}$. One should also  keep in mind, that fission 
observables are quite sensitive to pairing correlations owing to the 
strong dependence of the collective inertias on the inverse of the 
square of the pairing gap 
\cite{Robledo-Giulliani,proportional-1,proportional-2,Rayner-UPRC-2014,Rayner-UEPJA-2014,Rayner-RaEPJA-2016}.

Within the EDF framework, it is customary to describe the transition 
from a single even-even system to its fragments in terms of several 
deformation parameters 
\cite{Bjor,Specht,Krappe,Baran-Kowal-others-review2015,Schunck-Robledo-review}. 
Such microscopic (constrained) EDF calculations already represent a 
highly demanding computational task. They  are usually carried out 
using mean-field approximations  based on nonrelativistic Gogny 
\cite{Rayner-UPRC-2014,Rayner-UEPJA-2014,Rayner-RaEPJA-2016,gogny-d1s,Delaroche-2006,Dubray,Younes-fission,Warda-Egido-Robledo-Pomorski-2002,Warda-Egido-2012,Action-Rayner}, 
Skyrme 
\cite{Baran-Kowal-others-review2015,Schunck-Robledo-review,UNEDF1,Mcdonell-2,Erler2012,Baran-SF-2012,Baran-1981} 
and more recently Barcelona-Catania-Paris-Madrid (BCPM)  
\cite{BCPM,Robledo-Giulliani} as well as relativistic 
\cite{Abusara-2010,Abu-2012-bheights,RMF-LU-2012,Kara-RMF,Afa-arxiv} 
EDFs. What further complicates the EDF description of fission in 
odd-mass nuclei is the need to consider, for each intrinsic 
configuration along the fission path, blocked  one-quasiparticle wave 
functions \cite{rs}. The blocked wave functions break time-reversal 
invariance which requires the evaluation of time-odd fields. Typically, 
this requirement represents a factor of two more computing time in the 
solution of the Hartree-Fock-Bogoliubov (HFB) equations as compared to 
the even-even case where the time-odd fields are zero by construction. 
On the other hand, due to the (nonlinear) self-consistent character of 
the HFB equations, there is no guarantee to obtain the lowest energy 
solution by blocking the quasiparticle with the lowest 
one-quasiparticle excitation energy. Therefore, the blocking procedure 
has to be repeated for several low-lying one-quasiparticle states. 
 
From what has already  been mentioned above, it is clear that an 
approximation is required in order to afford the  computational cost of 
the  microscopic EDF description of fission in odd-mass nuclei. Within 
this context, the Equal Filling Approximation (EFA) represents a 
reasonable  starting point  still preserving  time-reversal invariance 
\cite{EFA-jsut,EFA-Bonneau,EFA-Rayner-1,EFA-Rayner-2,EFA-Rayner-3,EFA-Rayner-4,EFA-Decharge}. 
The EFA has been proven to be a fully variational method by recasting 
it in terms of the language of quantum statistical mechanics and the 
introduction of a statistical ensemble where one-quasiparticle 
configurations and their time-reversed companions are present with 
equal probability \cite{EFA-jsut}. One advantage of this point of view 
is that the results of finite temperature HFB and its extensions to 
compute collective inertias in the framework of  the Adiabatic Time 
Dependent (ATD) method \cite{ATDHFB-T-1,ATDHFB-T-2,ATDHFB-T-3} can be 
used to compute the collective inertias  needed to obtain the 
t$_\mathrm{SF}$ values for odd-mass nuclei.

Nuclei in the No (Z=102) region have attracted considerable attention 
\cite{No-REGION} thanks  to the progress made in  the production of 
super-heavy elements (see, for example, \cite{JULIN-SHE} and references 
therein) but also  because of the  rich spectroscopic data obtained for 
them \cite{SPEC-HERZ}. In this paper, we mainly focus on the HFB-EFA 
description of the fission properties in odd-mass nobelium isotopes. To 
the best of our knowledge our EDF calculations are the first of their 
kind reported in the literature. For the sake of completeness, we also 
discuss results for even-even No nuclei. We have studied the sample of 
nuclei $^{250-260}$No. The odd-mass isotopes are considered within the 
(constrained) HFB-EFA 
\cite{EFA-jsut,EFA-Bonneau,EFA-Rayner-1,EFA-Rayner-2,EFA-Rayner-3,EFA-Rayner-4,EFA-Decharge} 
while for the even-even ones calculations have been carried out using 
the standard (constrained) HFB framework 
\cite{Rayner-UPRC-2014,Rayner-UEPJA-2014,Rayner-RaEPJA-2016,rs}.

%
%

\begin{figure}
\includegraphics[width=0.46\textwidth]{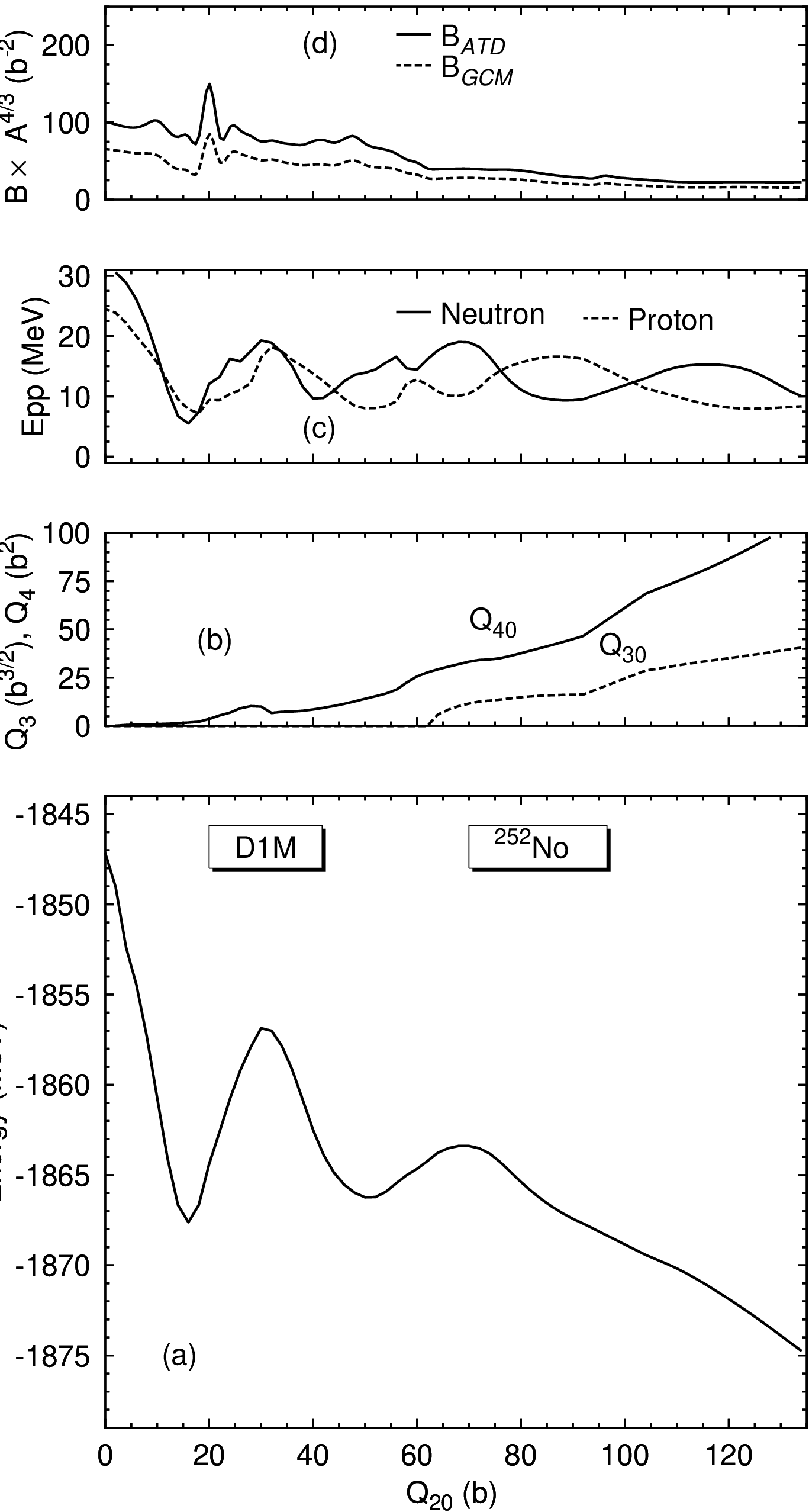}
\caption{
The HFB plus the zero-point rotational energies, are plotted in panel 
(a) as functions of the quadrupole moment $Q_{20}$ for the nucleus 
$^{252}$No. The octupole $Q_{30}$ and hexadecupole $Q_{40}$ moments are 
given in panel (b). The pairing interaction energy $E_{pp}=1/2\mathrm{Tr} \Delta \kappa^{*}$ is depicted in 
panel (c) for protons (dashed lines) and neutrons (full lines). The 
collective GCM (dashed lines) and ATD (full lines) masses are plotted 
in panel (d). For more details, see the main text. 
} \label{peda-252No} 
\end{figure}

Our HFB and/or HFB-EFA calculations are  based on the  parametrization 
D1M \cite{gogny-d1m} of the Gogny-EDF \cite{gogny}. In  previous 
studies \cite{Rayner-UPRC-2014,Rayner-UEPJA-2014,Rayner-RaEPJA-2016}, 
we have performed fission calculations for neutron-rich Ra, U and Pu 
nuclei as well as for a sample of heavy and super-heavy nuclei for 
which experimental data exist. The comparison between the D1S 
\cite{gogny-d1s}, D1N \cite{gogny-d1n} and D1M \cite{gogny-d1m} 
Gogny-EDFs, with available experimental data and other theoretical 
results, reveals that the parameter set D1M represents a reasonable 
starting point to describe fission in heavy and super-heavy nuclei. 
This is quite satisfying as the Gogny-D1M EDF does a much better job to 
reproduce nuclear masses \cite{gogny-d1m} and, at the same time, seems 
to keep essentially the same predictive power of the well tested D1S 
parametrization 
\cite{gogny-d1m,PRCQ2Q3-2012,Robledo-Rayner-JPG-2012,PTpaper-Rayner,Rayner-Robledo-JPG-2009,EFA-Rayner-3,EFA-Rayner-4} 
to reproduce  low-energy nuclear structure data. However, more work is 
needed in order to substantiate this conclusion, especially in the case 
of fission where the D1M parametrization has been scarcely used. In 
particular, in this study we explore, for the first time, the ability 
of the Gogny-D1M EDF framework to capture basic fission properties 
along the No isotopic chain including not only even-even but also 
odd-mass nuclei.

The paper is organized as follows. In Sec.~\ref{Theory}, we briefly 
outline the  HFB-EFA 
\cite{EFA-jsut,EFA-Rayner-1,EFA-Rayner-2,EFA-Rayner-3,EFA-Rayner-4,EFA-Decharge}. 
For details on the HFB framework for even-even nuclei the reader is 
referred to 
\cite{Rayner-UPRC-2014,Rayner-UEPJA-2014,Rayner-RaEPJA-2016,rs}. The 
results of our calculations are presented in Sec.~\ref{results} where 
we discuss the fission paths, ground state quantum numbers and 
deformations, first and second barrier heights, fission isomers, 
spontaneous fission and $\alpha$-decay half-lives  for  $^{250-260}$No 
and compare, whenever possible, with the available experimental data 
\cite{EXP-NO,Mass-Table-W}. In particular, we illustrate our methodology
for both  $^{252}$No and 
$^{253}$No in Secs.~\ref{example-252No} and 
\ref{example-253No} while the  systematic 
of the fission paths and spontaneous fission half-lives 
in $^{250-260}$No is presented in Secs.~\ref{systematics-paths} and 
\ref{systematics-TSF}. Let us stress that in this study, we 
implicitly assume that the properties of the considered No isotopes are 
determined by general features of the  Gogny-D1M EDF and therefore, no 
fine tuning of its parameters has been done. Finally, Sec. 
\ref{Coclusions} is devoted to the concluding remarks and work 
perspectives.

%
%

\begin{figure}
\includegraphics[width=0.46\textwidth]{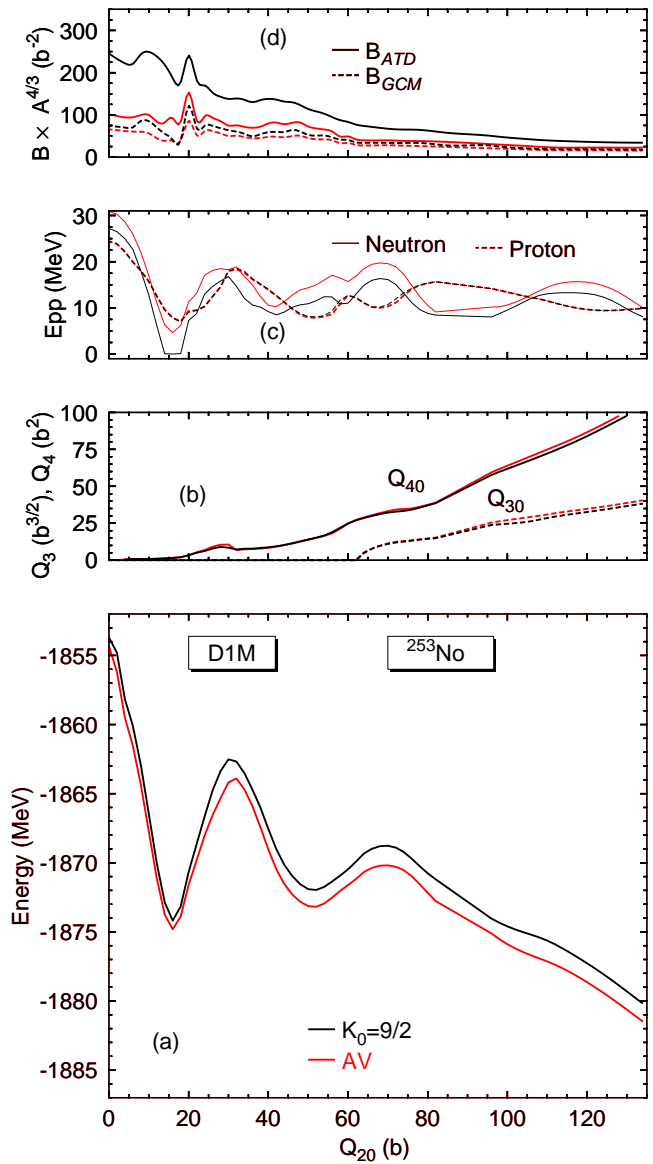}
\caption{ 
(Color online) The K$_{0}$ = 9/2  HFB-EFA plus the zero-point 
rotational energies, are plotted in panel (a) as functions of the 
quadrupole moment $Q_{20}$ for the nucleus $^{253}$No. The octupole 
$Q_{30}$ and hexadecupole $Q_{40}$ moments are given in panel (b). The 
pairing interaction energies are depicted in panel (c) for protons 
(dashed lines) and neutrons (full lines). The collective GCM (dashed 
lines) and ATD (full lines) masses are plotted in panel (d). Results 
corresponding to  "average" (AV)  HFB calculations for $^{253}$No have 
also been included in each panel (red). For more details, see the main 
text.
}
\label{peda-1} 
\end{figure}

%
%

\section{Theoretical framework}
\label{Theory}

In this section, we briefly  outline the  (constrained) HFB-EFA 
\cite{EFA-jsut,EFA-Rayner-1,EFA-Rayner-2,EFA-Rayner-3,EFA-Rayner-4,EFA-Decharge} 
used to compute the fission paths for $^{251,253,255,257,259}$No. 
Calculations for $^{250,252,254,256,258,260}$No have been carried out 
within the  (constrained) HFB approach and, for details, the reader is 
referred to 
\cite{Rayner-UPRC-2014,Rayner-UEPJA-2014,Rayner-RaEPJA-2016,rs}.

The description of odd-mass systems requires, blocked (product) 
one-quasiparticle wave functions $|{\Psi}_{\mu} \rangle= 
\hat{\alpha}_{\mu}^{\dagger} |{\Psi} \rangle$, which are given in terms 
of a fully paired reference vacuum  $|{\Psi} \rangle$ and the 
quasiparticle creator \cite{rs} $\hat{\alpha}_{\mu}^{\dagger}$ for the 
state $\mu$ to be blocked. The states $|{\Psi}_{\mu} \rangle$, as well 
as their density matrix and pairing tensor break the time-reversal 
invariance and so do the Hartree-Fock and pairing fields \cite{rs} 
associated with them. Furthermore, the density matrix and pairing 
tensor are no longer invariant under unitary transformations of the 
Bogoliubov amplitudes U and V \cite{rs} applied to the right and 
therefore, reorientation effects \cite{reorient-1,reorient-2} should 
also be taken into account in the solution of the mean-field equations. 
Therefore, in order to alleviate the numerical effort, we have resorted 
to the HFB-EFA within which one considers  the following density matrix 

%
%

\begin{figure*}
\includegraphics[width=1.00\textwidth]{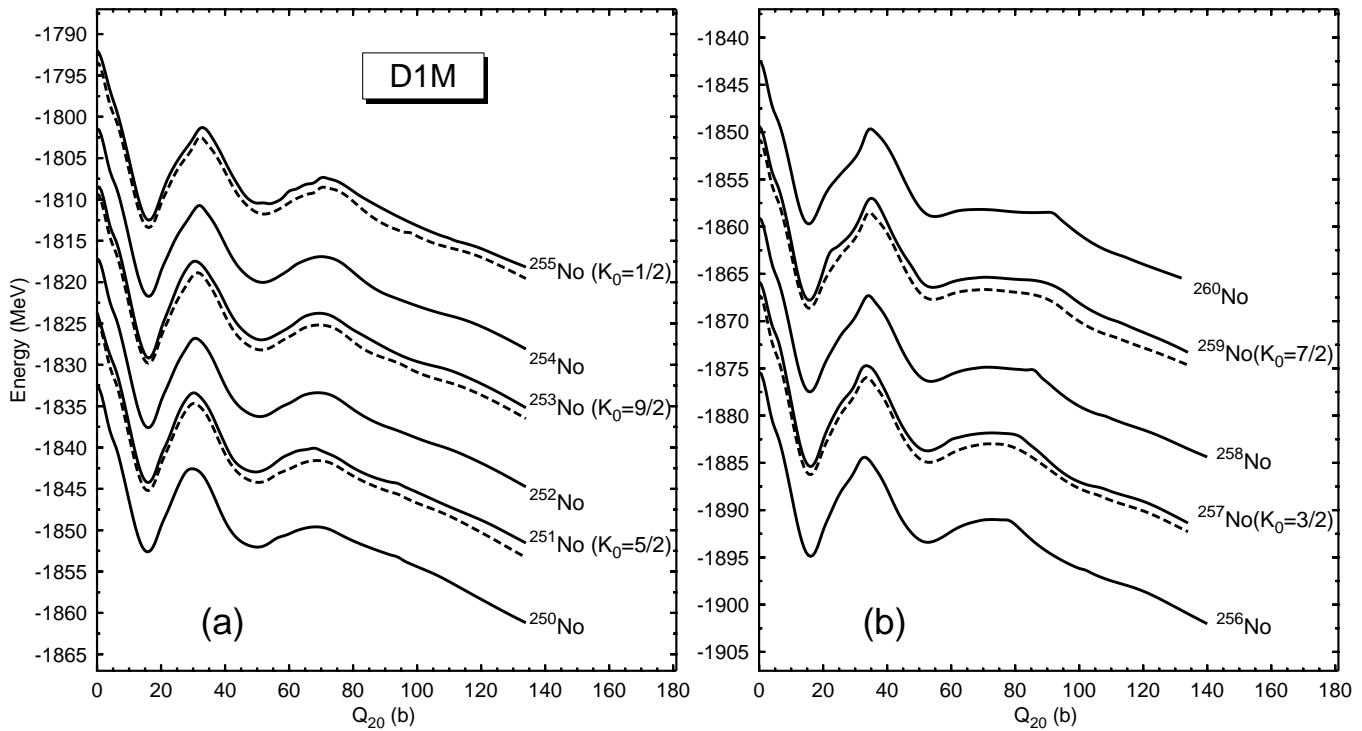}
\caption{
The ground state fission paths obtained for the isotopes $^{250-260}$No 
are plotted as functions of the quadrupole moment $Q_{20}$. Starting 
from the nucleus $^{251}$No ($^{257}$No ) in panel (a) [panel (b)], the 
curves have been shifted by 15 MeV in order to accommodate them in a 
single plot. For  $^{251,253,255,257,259}$No, the corresponding   
K$_{0}$ values are  included in the plot. The "average" (AV) fission 
paths for those odd-mass nuclei are also depicted with dashed lines. 
For more details, see the main text. 
}
\label{FissionBarriers-1} 
\end{figure*}

\begin{eqnarray} \label{rho-EFA}
{\rho}_{ij}^{(\mu,EFA)} &=& 
\frac{1}{2}
\left(
\langle {\Psi}_{\mu} |  \hat{c}_{j}^{\dagger} \hat{c}_{i}  | {\Psi}_{\mu} \rangle
+
\langle {\Psi}_{\overline{\mu}} |  \hat{c}_{j}^{\dagger} \hat{c}_{i}  | {\Psi}_{\overline{\mu}} \rangle
\right)
\nonumber\\
&=&
\left(V^{*}V^{T} \right)_{ij}
+\frac{1}{2}
\left(U_{i \mu} U_{j \mu}^{*} - V_{i \mu}^{*} V_{j \mu} \right)
\nonumber\\
&+&
\frac{1}{2}
\left(U_{i \overline{\mu}} U_{j \overline{\mu}}^{*} - V_{i \overline{\mu}}^{*} V_{j \overline{\mu}} \right)
\end{eqnarray}
and pairing tensor 

\begin{eqnarray} \label{kappa-EFA}
{\kappa}_{ij}^{(\mu,EFA)} &=& 
\frac{1}{2}
\left(
\langle {\Psi}_{\mu} |  \hat{c}_{j} \hat{c}_{i}  | {\Psi}_{\mu} \rangle
+
\langle {\Psi}_{\overline{\mu}} |  \hat{c}_{j} \hat{c}_{i}  | {\Psi}_{\overline{\mu}} \rangle
\right)
\nonumber\\
&=&
\left(V^{*}U^{T} \right)_{ij}
+\frac{1}{2}
\left(U_{i \mu} V_{j \mu}^{*} - V_{i \mu}^{*} U_{j \mu} \right)
\nonumber\\
&+&
\frac{1}{2}
\left(U_{i \overline{\mu}} V_{j \overline{\mu}}^{*} - V_{i \overline{\mu}}^{*} U_{j \overline{\mu}} \right)
\end{eqnarray}

Intuitively, ${\rho}_{ij}^{(\mu,EFA)}$ and ${\kappa}_{ij}^{(\mu,EFA)}$ 
correspond to an occupancy of 1/2 for the quasiparticle state $\mu$ and 
its Kramers' partner $\overline{\mu}$, thus preserving time-reversal 
invariance. This intuition can be substantiated by proving the 
existence of a statistical density matrix operator \cite{EFA-jsut} that 
produces exactly the same density matrix and pairing tensor. The help of 
Gaudin's theorem \cite{Gaudin-the,Gaudin-the1} is required to compute all the basic 
contractions in Eqs. (\ref{rho-EFA}) and (\ref{kappa-EFA}), as well as 
the mean values of any kind of operator within the HFB-EFA. The total 
HFB-EFA energy, which is the statistical average of the EDF with the 
special statistical density described above, can then be written in the 
usual form in terms  of ${\rho}_{ij}^{(\mu,EFA)}$ and 
${\kappa}_{ij}^{(\mu,EFA)}$ \cite{rs}. The Ritz-variational principle 
applied to the HFB-EFA energy leads to the HFB-EFA equations that can 
be solved using the standard gradient method 
\cite{Robledo-Bertsch2OGM}, with the subsequent simplification in the 
treatment of several constrains 
\cite{EFA-jsut,EFA-Rayner-1,EFA-Rayner-2,EFA-Rayner-3,EFA-Rayner-4}.

%
%

\begin{figure*}
\includegraphics[width=1.00\textwidth]{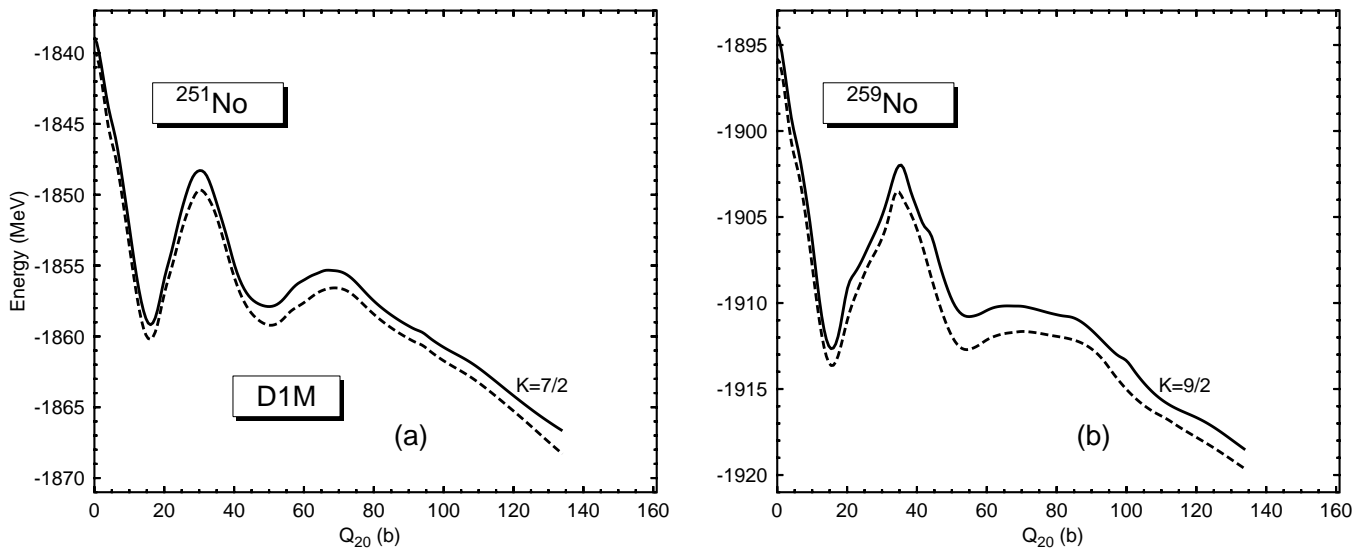}
\caption{
The K = 7/2 and 9/2 fission paths  obtained for the isotopes $^{251}$No 
and $^{259}$No are plotted as functions of the quadrupole moment 
$Q_{20}$ in panels (a) and (b), respectively. The "average" (AV) 
fission paths are also depicted with dashed lines. For more details, 
see the main text. 
}
\label{barriers-other-251-259No} 
\end{figure*}

Let us now turn our attention to the practical aspects of the  
methodology employed  in this work to obtain the ground state fission 
path in the case of $^{253}$No (see, Sec.~\ref{example-253No}) taken, as an illustrative example. The 
same methodology has been used for all the studied odd-mass No isotopes.


{\bf{Step 1}}) We have computed the fission path for $^{252}$No
(see, Sec.~\ref{example-252No})
 using 
the HFB approach with constrains on the axially symmetric quadrupole 
$\hat{Q}_{20}$ and octupole $\hat{Q}_{30}$ operators 
\cite{PRCQ2Q3-2012,Robledo-Rayner-JPG-2012}. Here, as well as in Steps 
2, 3 and 4 below, we have imposed an additional   constrain on the 
operator $\hat{Q}_{10}$, to avoid spurious center-of-mass effects. We 
have used a deformed  axially symmetric harmonic oscillator (HO)  basis 
consisting of states with J$_{z}$ quantum numbers up to 35/2 and up to 
26 quanta in the $z$-direction \cite{Rayner-UPRC-2014}. The oscillator 
lengths $b_{z}$ and $b_{\perp}$ for each value of 
$Q_{20} =  \langle \hat{Q}_{20} \rangle$ have been 
optimized as to minimize the energy. Zero-point quantum rotational 
$\Delta E_\mathrm{ROT}$ and vibrational $\Delta E_\mathrm{vib}$ 
energies have been added {\it{a posteriori}} to the HFB energies 
\cite{Rayner-UPRC-2014,Rayner-UEPJA-2014,Rayner-RaEPJA-2016}.

{\bf{Step 2}}) Using the HFB states (Step 1) as initial wave functions 
in the iterative mean-field procedure, we have computed  an "average"  
(AV) fission path for $^{253}$No (see, Fig.~\ref{peda-1}). In order to 
obtain the AV  HFB states, calculations have been carried out  as for 
an even-even nucleus but with a constrain on the mean value of the 
neutron number operator to be $\langle \hat{N} \rangle$=151. We have 
used the  single-particle basis with the (optimized)  HO lengths $b_{z}$ 
and $b_{\perp}$ resulting from Step 1. Zero-point quantum corrections 
have also been added to the mean-field energies 
\cite{Rayner-UPRC-2014,Rayner-UEPJA-2014,Rayner-RaEPJA-2016}.  
  
  
{\bf{Step 3}}) Using the HFB wave function corresponding to the 
absolute (normal deformed) minimum of the AV path (Step 2) in 
$^{253}$No as starting state, we have performed a set of HFB-EFA 
blocking calculations  to identify the  K=K$_{0}$ value corresponding 
to the  ground state. To this end, we have repeated the blocking 
procedure several times, using the same single-particle basis as in 
Step 2. We have  obtained three different solutions of the HFB-EFA 
equations for each of the K values from  K=1/2 up to 11/2. Larger K 
values have not been taken into account as the neutron single-particle 
levels corresponding to them are too far from the Fermi surfaces. In 
addition to K$_{0}$, the ground states obtained for all the odd-mass 
isotopes studied in this work  are characterized by their parity $\pi$. 
We will refer to them hereafter, as K$_{0}^{\pi}$ configurations.


{\bf{Step 4}}) Using the AV HFB states (Step 2) as starting wave 
functions, we have computed  the K$_{0}$ (i.e., ground state) fission 
path for $^{253}$No. Note that we are assuming that for an odd-mass 
nucleus with the ground state quantum numbers K$_{0}^{\pi}$, the 
spontaneous fission will take place in a configuration with the same 
K$_{0}$  (parity  can be broken in the fission process). For each 
(Q$_{20}$, Q$_{30}$)-configuration, we have considered the blocking of  
several  quasiparticle states  so as to obtain three different 
K$_{0}$-solutions of the HFB-EFA equations. From these three 
K$_{0}$-solutions,  the one with the lowest energy  is used to build 
the ground state fission path for $^{253}$No. In this step we use the 
same single-particle basis as in Step 2 and with the same oscillator 
lengths. 

Let us  mention, that in our Gogny-D1M mean-field calculations (Steps 1 
to 4) the two-body kinetic energy correction has been fully taken into 
account in the Ritz-variational procedure while for the Coulomb 
exchange term we have considered  the Slater approximation. The 
spin-orbit contribution to the pairing field has been neglected.

The computational cost to produce a plain potential energy surface 
requiring 75 constrained HFB calculations is of the order of five hours 
in a modern personal computer. If the oscillator lengths have to be 
optimized at each $Q_{20}$ value then the cost is a factor of 5 higher 
(Step 1). The computational cost of the EFA is in principle similar to 
the one of HFB as time odd-fields are not considered. However, in the 
EFA case several initial blocked configurations (three in our case) 
have to be considered. In addition, the number of iteration required to 
solve the EFA problem is typically two or three times larger than the 
one to solve HFB and therefore the typical cost of an EFA calculation 
is ten times higher than the one of a HFB calculation.

For the rotational  correction to the HFB-EFA energies, we have taken 
the expression $\Delta E_\mathrm{ROT}=\langle \Delta \hat{J}^{2} 
\rangle /  2 {{\cal{T}}}_{Y}$. However, both  $\langle \Delta 
\hat{J}^{2} \rangle$ and the Yocooz ${{\cal{T}}}_{Y}$ moment of inertia 
have been computed using the formulas for even-even nuclei 
\cite{RCORR-1,RCORR-2,RCORR-3}. The reason is that an approximate 
angular momentum projection  has not yet been carried out within the 
HFB-EFA framework. Work along these lines is in progress and will be 
reported elsewhere. On the other hand, the Adiabatic Time Dependent HFB 
(ATDHFB) framework has already been developed in the realm of quantum 
statistical mechanics \cite{ATDHFB-T-1,ATDHFB-T-2,ATDHFB-T-3} allowing 
its verbatim extension to the HFB-EFA via its statistical density 
matrix operator \cite{EFA-jsut}. In particular, the ATD  vibrational 
energy correction reads
\begin{eqnarray} \label{EVIB-ATDHFBEFA}
\Delta E_{vib, ATD} = \frac{1}{4} \frac{{\cal{M}}_{-2}}{{\cal{M}}_{-1}^{2}}
B_{ATD}^{-1}
\end{eqnarray}
For the evaluation of the collective mass $B_{ATD}$ in 
Eq. (\ref{EVIB-ATDHFBEFA}), we have resorted to the cranking 
approximation \cite{crankingAPPROX,Giannoni-1,Giannoni-2,Libert-1999}. 
In this case 
\begin{eqnarray} \label{ATDHFBEFA-CM}
B_{ATD} = \frac{1}{2} \frac{{\cal{M}}_{-3}}{{\cal{M}}_{-1}^{2}}
\end{eqnarray}
with moments 
\begin{eqnarray} \label{moments}
\left({\cal{M}}_{-n}\right)_{ij} = 
\sum_{\mu \nu} \frac{{\cal{Q}}_{i \mu \nu} {\cal{Q}}_{j \nu \mu}}
{\left( \overline{{\cal{E}}}_{\mu} - \overline{{\cal{E}}}_{\nu}\right)^{n}} 
\left(\overline{{\rho}}_{\mu} - \overline{{\rho}}_{\nu} \right)
\end{eqnarray}
and  
\begin{eqnarray} \label{QREPRE}
{\cal{Q}}_{i} =
\left(\begin{array} {cc}
Q_{i}^{11} & Q_{i}^{20}\\
-Q_{i}^{20 *} & -Q_{i}^{11 *} \\
\end{array}
\right)
\end{eqnarray}
In Eq. (\ref{QREPRE}), Q$_{i}^{11}$ and Q$_{i}^{20}$ are the 11 and 
20-components  of the considered operator in the quasiparticle 
representation \cite{rs}. The matrix $\overline{{\rho}}$ takes the form
\begin{eqnarray}
\overline{{\rho}} =
\left(\begin{array} {cc}
f & 0\\
0 & 1-f \\
\end{array}
\right)
\end{eqnarray}
where the f's are the quasiparticle HFB-EFA occupation factors (1/2 for 
the $\mu$ and $\overline{\mu}$ levels, respectively) while  
\begin{eqnarray}
\overline{{\cal{E}}} =
\left(\begin{array} {cc}
E & 0\\
0 & -E \\
\end{array}
\right)
\end{eqnarray}
where E$_{\mu}$ represents the quasiparticle energy. 

%
%

\begin{figure}
\includegraphics[width=0.47\textwidth]{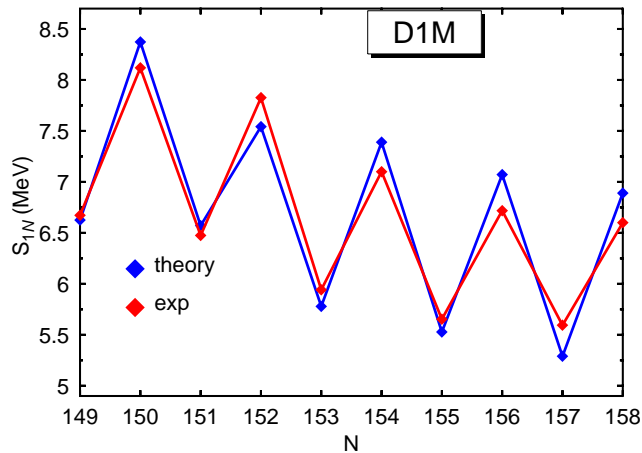}
\caption{
(Color online) The one-neutron S$_{1N}$ separation energies are plotted 
as functions of the neutron number. Experimental values are taken from 
\cite{Mass-Table-W}.
}
\label{S1N} 
\end{figure}

In the case of even-even nuclei, one usually considers the  GCM 
collective masses in addition to the ATD ones 
\cite{Rayner-UPRC-2014,Rayner-UEPJA-2014,Rayner-RaEPJA-2016}. Though 
the use of the former in the case of odd-mass systems lacks the 
theoretical justification available for the latter within the HFB-EFA, 
we have also considered their phenomenology in the present study in 
order to examine the impact of such a prescription  on the computed 
t$_{SF}$ values. To this end, for the collective GCM masses in  
odd-mass No isotopes we have  adopted the expression 
\begin{eqnarray} \label{Mass-GCM}
B_{GCM} = \frac{1}{2} \frac{{\cal{M}}_{-2}^{2}}{{\cal{M}}_{-1}^{3}}
\end{eqnarray}
and E$_\mathrm{vib, GCM}$ is given by Eq. (\ref{EVIB-ATDHFBEFA}) but 
replacing the ATD mass with the GCM one. The moments ${\cal{M}}_{-n}$  
are given by Eq. (\ref{moments}). For a thorough comparison of 
different forms of the collective masses in the framework of the 
Skyrme-EDF approximation, the reader is referred to \cite{Baran-TSF-2}. 
  
As a consequence of the cranking approximation, which essentially 
amounts to neglecting the residual interaction in the evaluation of 
collective masses, divergences might appear in the computation of the 
moments Eq. (\ref{moments}) due to level crossings in the denominator. A 
phenomenological way to deal with the problem is to replace those 
denominators  by $\sqrt{\left(E_{\mu}-E_{\nu} \right)^{2} + 
{\delta}^{2}}$, with $\delta$ being a regularization factor introduced 
to simulate the effect of the residual interaction \cite{Egido-reg}.

With all the required quantities at hand, we have computed the 
spontaneous fission half-life t$_\mathrm{SF}$ for the studied odd-mass 
nuclei using both the ATD and GCM schemes for the collective masses and 
zero-point vibrational corrections. For details regarding even-even 
isotopes the reader is referred to 
\cite{Rayner-UPRC-2014,Rayner-UEPJA-2014,Rayner-RaEPJA-2016}. Within 
the Wentzel-Kramers-Brillouin (WKB) formalism 
\cite{Baran-TSF-1,Baran-TSF-2} the t$_{SF}$ value (in seconds) is given by

\begin{eqnarray} \label{TSF-WKB}
\mathrm{t}_\mathrm{SF} = 2.86 \times 10^{-21} \times \left(1 + e^{2S} \right)
\end{eqnarray}
where the action S along the (minimal energy, one-dimensional projected)
fission path reads
\begin{eqnarray} \label{action}
S = \int_{a}^{b} dQ_{20} \sqrt{2B(Q_{20})\Big[V(Q_{20}) - (E_{GS}+E_{0}) \Big] }.
\end{eqnarray}
In this expression, the integration limits a and b are the classical 
turning points \cite{proportional-1} below the barrier corresponding to 
the energy E$_{GS}$+E$_{0}$. The collective potential V(Q$_{20}$) is 
given by the HFB-EFA energy corrected by the zero-point rotational 
$\Delta E_\mathrm{ROT}(Q_{20}$) and vibrational $\Delta E_\mathrm{vib}(Q_{20})$ 
energies. The parameter $E_{0}$  
accounts for the true ground state energy once the 
zero-point quadrupole fluctuations are considered. Although it is
not difficult to estimate its value using the curvature of the energy
around the ground state minimum and the values of the collective inertias 
\cite{Baran-SF-2012} we have followed the usual 
recipe \cite{Robledo-Giulliani,Warda-Egido-Robledo-Pomorski-2002}
of considering it as a free parameter that takes four different values
(i.e., $E_{0}$=0.5, 1.0, 1.5 and 2.0 MeV). In this way we can estimate 
its impact  on the predicted  spontaneous fission half-lives. Note, that
different $E_{0}$ values provide different classical turning points $a$ and 
$b$ in Eq. (\ref{action}) and therefore modify  the
final value of the action integral. As in previous studies 
\cite{Rayner-UPRC-2014,Rayner-UEPJA-2014,Rayner-RaEPJA-2016}, we have 
overlooked the E$_{0}$-dependence of the pre-factor in front of the 
exponential of the action in Eq. (\ref{TSF-WKB}) due to the large 
uncertainties in the estimation of the spontaneous fission half-lives 
arising from other sources.

Finally, we have computed the $\alpha$-decay half-lives t$_{\alpha}$ 
using  the Viola-Seaborg formula \cite{Viola-Seaborg}
\begin{eqnarray} \label{VSeaborg-new}
\log_{10} \mathrm{t}_{\alpha} =  \frac{AZ+B}{\sqrt{ {\cal{Q}}_{\alpha}}} + CZ+D + h_{log}
\end{eqnarray}
with the parameters A, B, C, D and h$_{log}$ given in \cite{TDong2005}. 
The ${\cal{Q}}_{\alpha}$ value is obtained from the calculated binding 
energies for No and Fm nuclei. 

%
%

\section{Discussion of the results}
\label{results}

In this section we discuss the results of our Gogny-D1M 
calculations. First, in Secs.~\ref{example-252No} and 
\ref{example-253No}, we consider the nuclei $^{252}$No and 
$^{253}$No as illustrative outcomes of our calculations. A similar analysis, as described
 below, has 
been carried out for all the studied even-even and odd-mass No isotopes. The systematic 
of the fission paths and spontaneous fission half-lives 
in $^{250-260}$No is presented in Secs.~\ref{systematics-paths} and 
\ref{systematics-TSF}, respectively.

\subsection{ The nucleus $^{252}$No}
\label{example-252No}

The HFB plus the zero-point rotational energies, are plotted in panel 
(a) of Fig.~\ref{peda-252No} as functions of the quadrupole moment for 
$^{252}$No. The zero point vibrational energies are not included in the 
plot, as they are rather constant as functions of Q$_{20}$. However, 
they are always included in the computation of the spontaneous fission 
and $\alpha$-decay half-lives. The octupole $Q_{30}$ and hexadecupole 
$Q_{40}$ moments are given in panel (b). The ground state for 
$^{252}$No is located at Q$_{20}$ = 16 b and is reflection symmetric. 
The fission isomer at Q$_{20}$ = 52 b lies 1.38 MeV above the ground 
state from which, it is separated by the inner  barrier (Q$_{20}$ = 30 
b) with the height of 10.75 MeV. As in the ground state case, the 
fission isomer is also reflection symmetric, a property that is shared 
for all the configurations belonging to the first barrier. However, as 
deduced from panel (b), octupole correlations play a role for 
quadrupole deformations Q$_{20} \ge$  62 b. Those correlations 
significantly affect the outer barrier (Q$_{20}$ = 70 b) whose height 
is 4.23 MeV.

%
%

\begin{figure}
\includegraphics[width=0.47\textwidth]{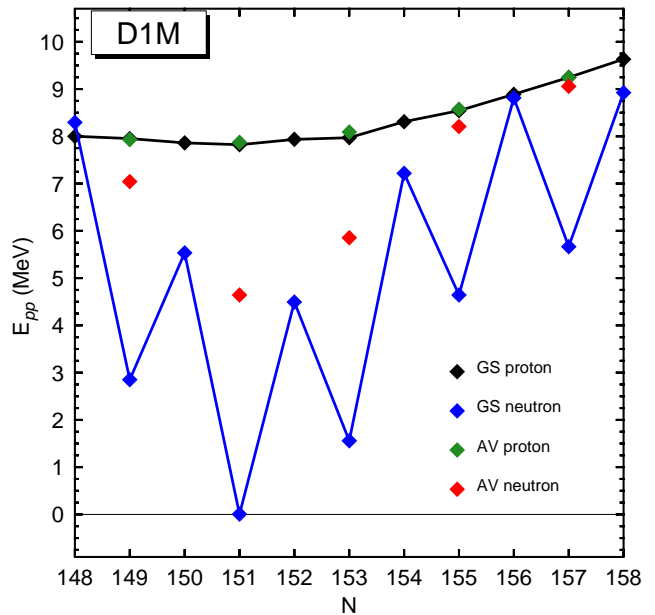}
\caption{(
Color online) The proton and neutron  pairing interaction energies 
E$_{pp}$ corresponding to the ground states (GS) of  $^{250-260}$No are 
plotted as functions of the neutron number N. The "average" (AV) 
E$_{pp}$  values obtained for odd-mass isotopes are also included in 
the plot.
}
\label{PAIR-vs-N-No} 
\end{figure}

The proton and neutron pairing interaction energies E$_{pp} =1/2 
\mathrm{Tr} \left(\Delta \kappa^{*} \right)$ are depicted in panel (c). 
The neutron E$_{pp}$ values exhibit minima (maxima) at Q$_{20}$ = 16, 
40 and 88b (30, 68 and 120 b) while for protons those minima (maxima) 
correspond to Q$_{20}$ = 18, 50 and 68b (30, 60 and 88 b). The minima 
of E$_{pp}$ are related to the minima of the potential energy surface 
(PES), a fact that can be understood in terms of level densities: 
minima of the PES correspond to low level density regions (Jahn-Teller 
effect) and the amount of pairing correlations is proportional to the 
level density around the Fermi level. On the other hand, maxima of the 
PES are associated to high level density regions implying stronger 
pairing correlations.

The collective GCM and ATD masses are plotted in panel (d). Their 
behavior is well correlated with the one of the pairing correlations in 
panel (c) and the inverse dependence of the collective mass with the 
square of the pairing gap \cite{proportional-1,proportional-2}. 
We also observe that the ATD masses are systematically larger than the GCM ones. 
For example, for Q$_{20}$ = 20 b we have obtained the ratio 
B$_{ATD}$/B$_{GCM}$ = 1.77. Such differences can have a strong impact 
on the predicted fission half-lives 
\cite{Robledo-Giulliani,Rayner-UPRC-2014,Rayner-UEPJA-2014,Rayner-RaEPJA-2016} 
and are the reason why both kinds of collective masses have been 
considered in the computation of the t$_\mathrm{SF}$ values. For 
example, for $^{252}$No and E$_{0}$ = 1.0 MeV we have obtained 
$\log_{10}$ t$_\mathrm{SF}$ = 8.7919 s within the ATD scheme while 
$\log_{10}$ t$_\mathrm{SF}$ = 6.9928 s within the GCM one. Another relevant 
quantity is the parameter E$_{0}$, because increasing it reduces the 
integration interval in the action and this 
leads to smaller t$_\mathrm{SF}$ values in either 
the ATD or GCM scheme. For example, once more in the case of $^{252}$No 
but with E$_{0}$ = 1.5 MeV, we have obtained $\log_{10}$ 
t$_\mathrm{SF}$ = 5.7252 s and $\log_{10}$ t$_\mathrm{SF}$ = 4.7378 s 
within the ATD and GCM approaches, respectively. In the computation of 
the fission half-lives, the wiggles in the collective masses have been 
softened by means of a three point filter 
\cite{Rayner-UPRC-2014,Rayner-UEPJA-2014,Rayner-RaEPJA-2016}.

\subsection{ The nucleus $^{253}$No}
\label{example-253No}

In Fig.~\ref{peda-1}, we have plotted similar curves as in 
Fig.~\ref{peda-252No} but for the fission path of $^{253}$No corresponding
to the K value of its ground state, i.e.,  K$_{0}=9/2$. 
Results corresponding to AV HFB calculations (see, Sec. 
\ref{Theory}) for this odd-mass nucleus have also been included in each 
panel for the sake of comparison. As can be seen from panel (a)  the 
K$_{0}$ = 9/2 energy, computed as the HFB-EFA  plus the rotational energy,  displays  as 
a
function of the quadrupole moment, a behavior reminiscent of the one 
found for $^{252}$No. The absolute minimum of the K$_{0}$ = 9/2 path 
appears at Q$_{20}$ = 16 b and corresponds to a parity-conserving 
configuration [see, panel (b)] with $\pi$ = -1, which agrees well with 
the experimental \cite{EXP-NO} K$^{\pi}$ = 9/2$^{-}$ ground state 
for this nucleus.  The  9/2$^{-}$ fission isomer at Q$_{20}$ = 52 b 
lies 2.22 MeV above the ground state from which, it is separated by the 
inner  barrier (Q$_{20}$ = 30 b) with the height of 11.67 MeV. The 
left-right symmetry is broken for Q$_{20}$ $\ge$ 62 b and the height of 
the outer  barrier (Q$_{20}$ = 70 b) turns out to be 5.42 MeV. All these
values are obtained with the K=9/2 configuration corresponding to the lowest
energy for each $Q_{20}$ value of the fission path. In this way, the 
number just given for the fission isomer might or might not correspond
to the lowest energy configuration with the deformation of the fission isomer.
In fact, in the case of $^{253}$No, the lowest energy configuration 
at Q$_{20}$ = 52 b corresponds to K=11/2
and  lies at an excitation energy of 2.04 MeV. 

On the other hand, the comparison between the EFA and AV paths in panel 
(a) of Fig.~\ref{peda-1}, reveals that the former is always higher than 
the latter and the difference is not constant with the quadrupole 
moment. The EFA inner and outer barriers (11.67 and 5.42 MeV)
 are higher than the AV ones 
(10.91 and 4.62 MeV). The same holds for the
EFA (2.22 MeV) and AV (1.63 MeV)
excitation energy of the fission isomer. This is a clear 
manifestation of the specialization energy effect partly due to 
following the lowest energy configuration with a fixed K$_{0}$ = 9/2  
value.  

Note, that the octupole $Q_{30}$ and hexadecupole $Q_{40}$ moments 
[panel (b)] corresponding to the AV and EFA fission  paths 
follow a rather similar pattern  revealing the minute impact of the
blocked configuration in the mass moments defining the nuclear shape.

As can be seen from panel (c) of Fig.~\ref{peda-1},  the HFB-EFA 
(black dashed line)
and AV (red dashed line)
proton E$_{pp}$ values can hardly be distinguished. Both the HFB-EFA 
and AV neutron pairing energies also display a similar pattern. 
However, the former are much lower than the latter as a consequence of 
the quenching of pairing correlations due to the unpaired neutron. A 
direct consequence of this quenching also reflects in the  enhancement 
of the HFB-EFA masses as compared to the AV ones regardless of the ATD 
and/or GCM scheme used [see, panel (d)]. From Fig.~\ref{peda-1}, one 
concludes that the effect of the unpaired neutron is to increase the
fission barrier heights and an enhancement of the HFB-EFA collective 
masses with respect to the AV ones. Both effects go in the direction of 
increasing the collective action Eq. (\ref{action}) and therefore, the 
t$_\mathrm{SF}$ values Eq.~(\ref{TSF-WKB}) increase in odd-mass nuclei 
as compared to their even-even neighbors.

Once more, one sees that the ATD masses are larger than the GCM ones 
leading, for example, to the values $\log_{10}$ t$_\mathrm{SF}$ = 24.5590 s and 
$\log_{10}$ t$_\mathrm{SF}$ = 15.3318 s within the ATD and GCM schemes for  
$^{253}$No and E$_{0}$ = 1.0 MeV. On the other hand, for E$_{0}$ = 1.5 
MeV we have obtained the values $\log_{10}$ t$_\mathrm{SF}$ = 20.8810 s and 
$\log_{10}$ t$_\mathrm{SF}$ = 12.9325 s within the ATD and GCM approaches, 
respectively. Due to these differences, and even when there is a lack 
of detailed theoretical justification for the use of GCM-like masses in 
the case of odd-mass systems, we have also considered both schemes in 
the computation of their spontaneous fission half-lives.

\subsection{Systematic of the fission paths in $^{250-260}$No}
\label{systematics-paths}

\begin{figure}
\includegraphics[width=0.46\textwidth]{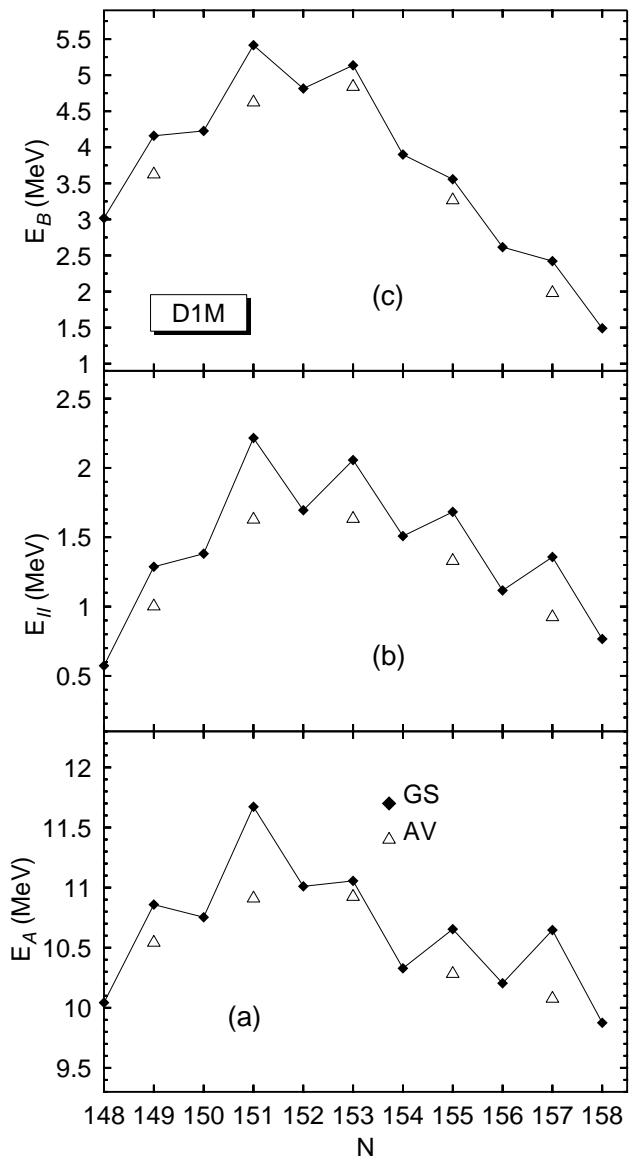}
\caption{ 
The inner barrier height E$_{A}$, excitation energy of the fission 
isomer E$_{II}$ and  outer barrier height  E$_{B}$ corresponding 
to the ground state (GS) fission paths shown in 
Fig.~\ref{FissionBarriers-1} are plotted in panels (a), (b) and (c) as 
functions of the neutron number. The "average" (AV)   values obtained 
for odd-mass isotopes are also included in the plot.
}
\label{BI-ISO-BII} 
\end{figure}

In Fig.~\ref{FissionBarriers-1}, we have summarized the ground state 
fission paths obtained for $^{250-260}$No as functions of the 
quadrupole moment Q$_{20}$. The  paths show 
two minima, the ground state one  at Q$_{20}$ = 16 b and the fission 
isomer around Q$_{20}$ = 52 b.  For all the studied nuclei, our 
Gogny-D1M calculations predict the super-deformed minimum to be above 
the normal deformed one though  its excitation energy  decreases with 
increasing neutron number [see, panel (b) of Fig.~\ref{BI-ISO-BII}]. On 
the other hand, the top of the inner and outer barriers is located at  
Q$_{20}$ $\approx$ 30 b and Q$_{20}$ $\approx$ 70 b, respectively. For 
all the paths shown in the figure, the outer barriers belong to a 
parity-breaking sector that decreases their heights as compared with a 
reflection-symmetric situation. The specialization energy effect 
becomes apparent from the comparison between the ground state and AV 
fission paths in the case of odd-mass nuclei.

%
%

In good agreement with the experimental data \cite{EXP-NO}, our 
calculations predict 9/2$^{-}$, 1/2$^{+}$ and 3/2$^{+}$ ground states 
for $^{253}$No, $^{255}$No and $^{257}$No. This is not the case for the 
nuclei $^{251}$No and $^{259}$No where we have obtained 5/2$^{+}$  and 
7/2$^{+}$ ground states  while the experimental ones correspond to 
7/2$^{+}$ and 9/2$^{+}$, respectively. However, for those nuclei we 
have  found 7/2$^{+}$ and 9/2$^{+}$ one-quasiparticle configurations 
located inside the ground state well whose excitation energies are only 
65.55 and 149.62 keV, respectively. Therefore, we have also explored 
the K = 7/2 and 9/2  fission paths in the case of $^{251}$No and 
$^{259}$No. They are plotted in Fig.~\ref{barriers-other-251-259No} as 
functions of Q$_{20}$. As can be seen, the structure of those paths and 
the associated specialization energy effect resemble the ones in 
Fig.~\ref{FissionBarriers-1} (see also, the discussion below). 

%
%

\begin{figure*}
\includegraphics[width=1.\textwidth]{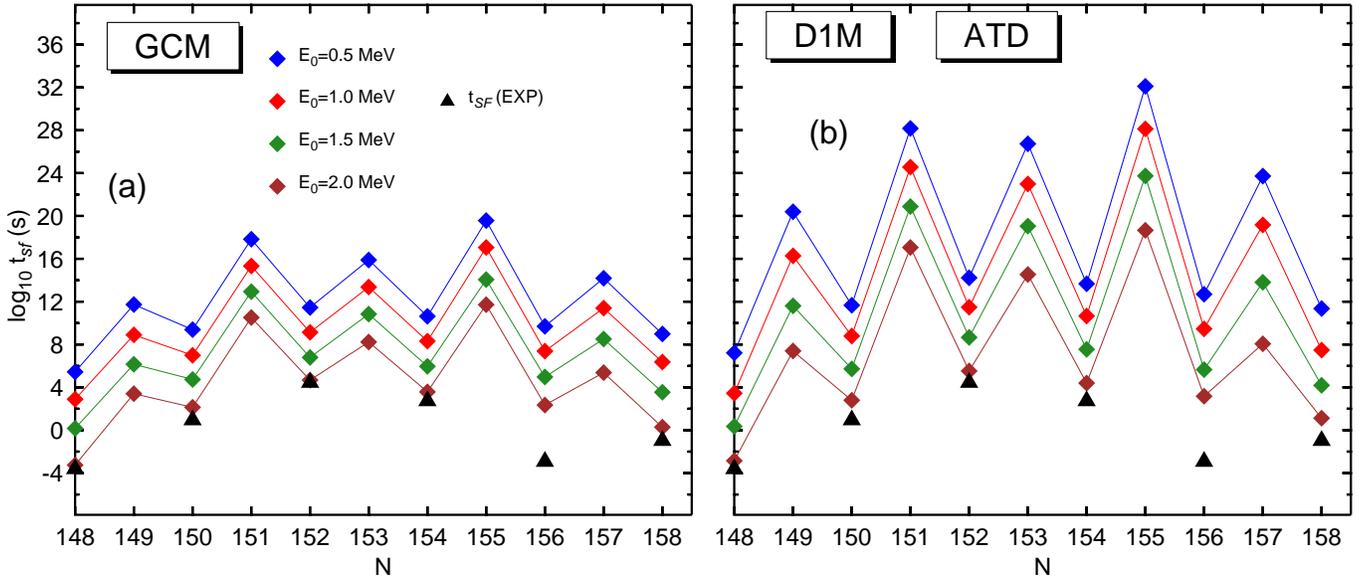}
\caption{ 
(Color online) The spontaneous fission half-lives t$_\mathrm{SF}$, predicted 
within the GCM [panel (a)] and ATD [panel (b)] schemes, for the 
isotopes $^{250-260}$No are depicted as functions of the neutron number 
N. For each isotope, calculations have been carried out with E$_{0}$ = 
0.5, 1.0, 1.5 and 2.0 MeV. The experimental  t$_\mathrm{SF}$ values for 
even-even isotopes are taken from \cite{Holden-paper}. For 
more details, see the main text.
}
\label{tsf-figure} 
\end{figure*}

Using the ground state energies obtained  for $^{250-260}$No, we have 
computed the one-neutron S$_{1N}$ separation energies  shown in 
Fig.~\ref{S1N}. Though our calculations capture the experimental   
\cite{Mass-Table-W}  odd-even staggering, the predicted amplitudes  are 
larger than the experimental ones,  especially for the heavier 
isotopes. This could be a consequence of an inappropriate pairing 
strength in the regions under study or the neglect of many-body effects
like particle-vibration coupling or the restoration of the broken
U(1)
 particle
number symmetry. Another source of uncertainty is the neglect of time-odd
fields in the EFA used in the present calculation that can contribute
up to a few hundred keV to the binding energy of odd nuclei \cite{Rob.12}.

The proton and neutron  pairing interaction energies E$_{pp}$ 
corresponding to the ground states  of  $^{250-260}$No are plotted in 
Fig.~\ref{PAIR-vs-N-No} as functions of the neutron number N. The AV 
E$_{pp}$  values obtained for $^{251,253,255,257,259}$No are also 
included in the plot. For protons the  pairing energies exhibit a 
smooth increase with increasing neutron number.  As expected, they are  
rather similar to the AV ones for $^{251,253,255,257,259}$No. However, 
in the case of neutrons, the quenching of the pairing correlations 
induced by blocking is very severe.  We have obtained the ratios 
E$_{pp,GS}$/E$_{pp,AV}$ = 0.40, 0.21, 0.26, 0.56 and 0.62 for 
$^{251,253,255,257,259}$No. An alternative measure of how effective is 
the quenching of the neutron pairing correlations is provided by the 
ratio $\langle \Delta \hat{N}^{2} \rangle_{GS}$/$\langle \Delta 
\hat{N}^{2} \rangle_{AV}$ = 0.528, 0.08, 0.332, 0.79 and 0.821 for the 
same nuclei. For the K$^{\pi}$ = 7/2$^{+}$ (9/2$^{+}$) normal deformed 
minimum in $^{251}$No ($^{259}$No) we have obtained the ratios 
E$_{pp,GS}$/E$_{pp,AV}$ = 0.29 (0.63) and $\langle \Delta \hat{N}^{2} 
\rangle_{GS}$/$\langle \Delta \hat{N}^{2} \rangle_{AV}$ = 0.422 
(0.824). These results indicate that for the neutron pairing 
correlations the quenching induced by the self-consistent blocking 
depends on the considered one-quasiparticle configuration. They also 
show that, at least for some of the studied odd-mass isotopes (for 
example, $^{253}$No)  one is dealing with a weak  pairing  regime. 
Therefore, a more realistic account of the fission process in those 
systems would require a dynamical description of the neutron pairing 
correlations as well as  their coupling to the relevant deformation 
parameters \cite{Action-Rayner}.  

%
%

\begin{table}
\caption{The values of log$_{10}$ t$_\mathrm{SF}$ (in s) obtained for the 
K=7/2 and 5/2 fission paths in $^{251}$No are given as functions of the 
parameter E$_{0}$ (in MeV). The same quantities are also given for the 
K=9/2 and 7/2 fission paths in $^{259}$No. For details, see the main text.
} 
\begin{tabular}{ccccccc}
\hline\hline
Nucleus     & K       & Scheme          & \multicolumn{4}{c}{  E$_{0}$}                                                     \\
            &         &                 &         0.5 MeV   &           1.0 MeV &           1.5 MeV  &           2.0 MeV    \\
\hline\hline
$^{251}$No  & 7/2     &  GCM            &  11.3558          &  8.7824           &   6.2380           &    3.5589            \\ 
$^{251}$No  & 7/2     &  ATD            &  19.9848          &  16.0185          &   11.3774          &    7.0114            \\   
$^{251}$No  & 5/2     &  GCM            &  11.7274          &  8.8949           &   6.1555           &    3.4046            \\
$^{251}$No  & 5/2     &  ATD            &  20.3827          &  16.2834          &   11.6087          &    7.3890            \\
\\
$^{259}$No  & 9/2     &  GCM            &  13.2903          &  10.6305           &   7.9765           &    5.3722           \\
$^{259}$No  & 9/2     &  ATD            &  22.9660          &  18.8139           &   14.2703          &    8.0629          \\
$^{259}$No  & 7/2     &  GCM            &  14.1852          &  11.3998           &   8.5118           &    5.1894           \\
$^{259}$No  & 7/2     &  ATD            &  23.7159          &  19.1671           &   13.8127           &    8.0677           \\
\hline\hline
\end{tabular}
\label{other-tsf-251-259}
\end{table}

The inner barrier heights E$_{A}$, excitation energies of the fission 
isomers E$_{II}$ and  outer barrier heights E$_{B}$ 
corresponding to the ground state fission paths shown in 
Fig.~\ref{FissionBarriers-1} are displayed in panels (a), (b) and (c) 
of Fig.~\ref{BI-ISO-BII} as functions of the neutron number N. The 
corresponding AV quantities for $^{251,253,255,257,259}$No are also 
depicted. The three quantities exhibit pronounced odd-even as well as 
specialization energy effects. In our calculations the largest 
E$_{A}$, E$_{II}$ and E$_{B}$ values are obtained for the 
nucleus $^{253}$No (N = 151). Moreover, for the K = 7/2 (9/2)  fission 
path in $^{251}$No ($^{259}$No) we have obtained E$_{A}$ = 10.88 
(10.54) MeV, E$_{II}$ = 1.28 (1.86) MeV and E$_{B}$ = 3.83 
(2.46) MeV. These values should be compared with the ones [i.e., 10.86 
(10.65) MeV, 1.29 (1.36) MeV and 4.16 (2.42) MeV] shown in the figure 
which correspond to the ground state paths predicted for  
those nuclei. Let us mention, that we are aware of the reduction, by a 
few MeV,  of the inner barrier heights once triaxiality is taken into 
account \cite{Rayner-UPRC-2014}. In the case of odd-mass nuclei, the 
polarization effects induced by the unpaired nucleon might also lead to 
triaxial solutions. Though we have kept axial symmetry along the 
fission paths of the studied No isotopes, we have also corroborated the 
already mentioned reduction in the case of the even-even nuclei 
$^{254}$No and $^{256}$No. However, as such a reduction comes together 
with an increase in the collective inertia 
\cite{Baran-1981,Bender-1998} that tends to compensate in the final 
value of the action, the role of triaxiality is very limited and has 
not been taken into account in the computation of the corresponding 
t$_\mathrm{SF}$ values.

\subsection{Systematic of the spontaneous fission half-lives in $^{250-260}$No}
\label{systematics-TSF}

The ground state spontaneous fission half-lives t$_\mathrm{SF}$, 
obtained within the GCM  and ATD schemes, for the isotopes 
$^{250-260}$No are depicted, as functions of the neutron number N, in 
panels (a) and (b) of Fig.~\ref{tsf-figure}, respectively. In the  
panels we have included the available experimental data for even-even 
No isotopes \cite{Holden-paper}. Those data are more scarce for 
odd-mass nuclei for which only some recommended lower bound values (t$_\mathrm{SF}$ 
$\ge$ 10 s for $^{251}$No,  t$_\mathrm{SF}$ $\ge$ 28 min for $^{257}$No 
and t$_\mathrm{SF}$ $>$ 10 h for $^{259}$No) can be found in the 
literature \cite{Holden-paper}. However, in spite of being only lower 
bounds, those data already reveal an increase in the spontaneous 
fission half-lives of the odd-mass isotopes. In our calculations, for 
each of the studied isotopes, we have considered four values of E$_{0}$ 
(i.e., E$_{0}$ = 0.5, 1.0, 1.5 and 2.0 MeV). As can be seen from the 
figure, increasing this parameter leads to a decrease in the predicted 
spontaneous fission half-lives as well as to an improvement (exception 
made of $^{258}$No) of the agreement with the experiment in the case of 
even-even nuclei . The predicted ATD t$_{SF}$ values tend to be larger 
than the GCM ones, with the differences being more pronounced in the 
case of odd-mass nuclei. For example, for $^{254}$No ($^{257}$No) and  
E$_{0}$ = 1.0 MeV we have obtained $\log_{10}$ t$_\mathrm{SF}$ = 9.1373 
s ($\log_{10}$ t$_\mathrm{SF}$ = 17.0551 s) within the GCM scheme while 
the corresponding value within the ATD approach is 
$\log_{10}$ t$_\mathrm{SF}$ = 11.4739 s ($\log_{10}$ t$_\mathrm{SF}$ = 
28.1227 s) [see also, 
Table~\ref{other-tsf-251-259}]. Though there exists a strong variance 
of the theoretical fission rates with respect to the details involved 
in their computation, the same trend is always observed, i.e., 
regardless of the employed scheme and/or E$_{0}$ value the fission 
half-lives exhibit a pronounced odd-even staggering. The amplitude of
the staggering (i.e., the difference between the spontaneous fission
lifetime of an odd-mass isotope and the one obtained by averaging the 
values of the two neighboring isotopes) is rather insensitive to the E$_{0}$ value considered but
it depends strongly both on neutron number and on the kind of collective inertia considered. 
The amplitude varies from the four orders of magnitude observed for the
N=149 and N=157 isotopes in the calculation with the GCM inertias to the 16
orders of magnitude for the No isotope with N=155 in the ATD calculation. The amplitude
of the staggering is larger in the ATD than in the GCM case (up to a 
factor 1.6) as can be expected from the values of the corresponding inertias.
The dependence with neutron number shows a seesaw pattern with large
values at N=151 and N=155.  On the other hand,
the amplitude of the staggering seems to be an interesting quantity to be compared with
experimental data given its insensitivity to the E$_{0}$ values. 
Similar 
features to the ones just discussed also emerge by  comparing the  spontaneous fission half-lives, 
given in Table~\ref{other-tsf-251-259}, for the   5/2 and 7/2 fission 
paths in $^{251}$No  as well as the 9/2 and 7/2 ones in $^{259}$No 
(see, Figs.~\ref{FissionBarriers-1} and 
\ref{barriers-other-251-259No}).

%
%

\begin{figure*}
\includegraphics[width=1.\textwidth]{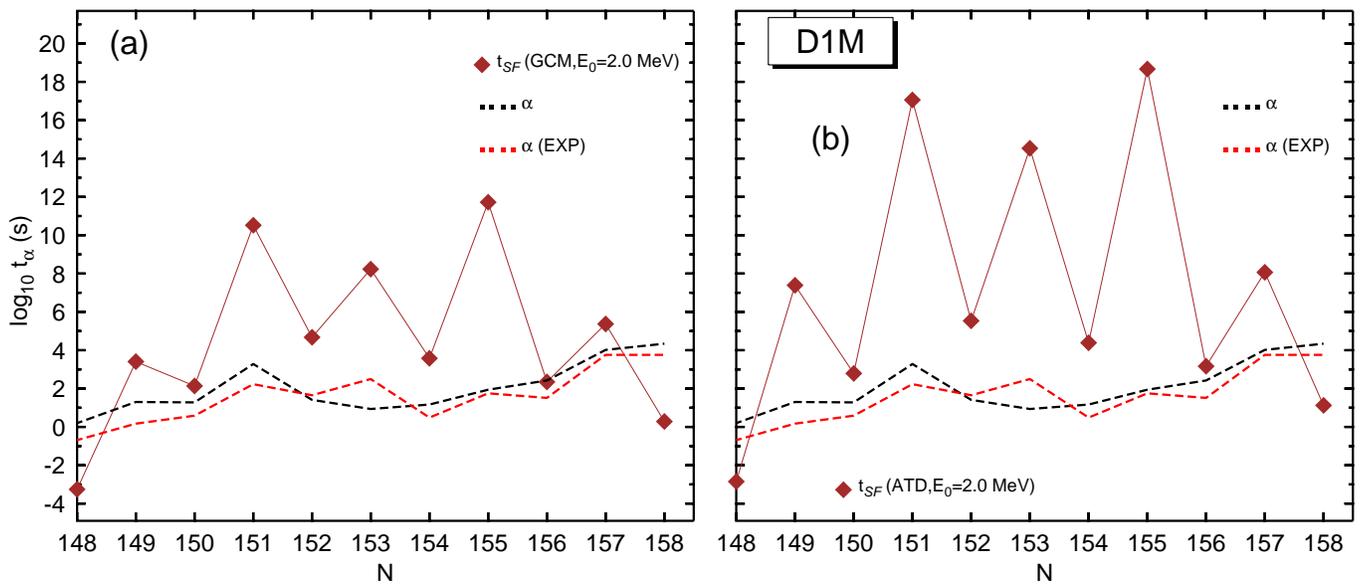}
\caption{ 
(Color online) The 
theoretical and experimental \cite{Mass-Table-W} $\alpha$-decay 
half-lives are plotted as functions of the neutron number 
N. They are compared in panel (a) [panel (b)] with the spontaneous 
fission half-lives obtained within the GCM (ATD) scheme with 
E$_{0}$ = 2.0 MeV. For more details, see the main text.
}
\label{talpha-figure} 
\end{figure*}

In Fig.~\ref{talpha-figure}, we have  plotted  the $\alpha$-decay 
half-lives computed with a parametrization \cite{TDong2005} of the 
Viola-Seaborg formula \cite{Viola-Seaborg}. To this end, we have used 
the binding energies obtained for the corresponding nobelium and 
fermium isotopes. The $\alpha$-decay half-lives obtained in our 
Gogny-D1M calculations compare well with the experimental ones 
\cite{Mass-Table-W}  as could be expected from
the good performance of the Gogny-D1M EDF in reproducing experimental binding
energies and therefore in reproducing the $\mathcal{Q}_{\alpha}$ values
entering the Viola-Seaborg formula. In panel (a), we have also included 
the spontaneous fission half-lives predicted within the GCM scheme 
(E$_{0}$ = 2.0 MeV). From the comparison between the t$_{\alpha}$ and 
GCM  t$_\mathrm{SF}$ values we 
conclude that $\alpha$-decay is  the dominant decay channel
for the odd-mass No isotopes considered even for $^{259}$No that lies
in between two even-even nuclei where spontaneous fission dominates. The same 
conclusion can be obtained from the comparison  with the 
corresponding ATD fission lifetimes in panel (b).

%
%

\section{Conclusions}
\label{Coclusions}

In this paper, we have considered the  fission properties  of a set of 
nobelium isotopes including even-even and odd-mass nuclei. The selected 
set comprises $^{250-260}$No as a representative sample. We 
have resorted to the constrained Gogny-D1M mean-field approximation as 
our main theoretical tool. In particular, the even-even systems have 
been studied within the standard HFB framework while for the odd-mass 
isotopes we have used the HFB-EFA in order to reduce the already 
substantial computational effort. Besides the proton $\hat{Z}$ and 
neutron $\hat{N}$  number operators, we have employed constrains on the 
axially symmetric quadrupole $\hat{Q}_{20}$, octupole  $\hat{Q}_{30}$ 
and $\hat{Q}_{10}$ operators. We have presented a detailed description 
of the (blocking) methodology used to obtain the (least energy) fission 
paths in the studied odd-mass isotopes while for the even-even ones 
calculations have been carried out along the lines discussed in 
previous fission studies 
\cite{Rayner-UPRC-2014,Rayner-UEPJA-2014,Rayner-RaEPJA-2016}. 
Zero-point rotational and vibrational corrections have always been 
added to the HFB and/or HFB-EFA energies {\it{a posteriori}}. The 
rotational correction has been found in terms of the Yoccoz moment of 
inertia while both the GCM and ATD schemes have been used to compute 
the collective masses and vibrational corrections. All the required 
mean-field building blocks have then been used to compute the GCM 
and/or ATD ground state spontaneous fission half-lives t$_\mathrm{SF}$ 
for $^{250-260}$No within the WKB formalism. The $\alpha$-decay 
half-lives t$_{\alpha}$ have been found,  using the binding energies 
obtained for the corresponding No and Fm nuclei, with the help of a 
 parametrization of the Viola-Seaborg formula. 

For both even-even and odd-mass No isotopes our Gogny-D1M calculations 
provide ground state fission paths with normal deformed and isomeric 
minima as well as inner and outer barriers whose quadrupole 
deformations, remain almost constant as functions of the neutron 
number. In all cases, the outer barriers belong to a 
parity-breaking sector that decreases their heights as compared to a 
reflection-symmetric situation. The inner and outer barrier heights as 
well as the isomer excitation energies display pronounced odd-even 
effects as functions of the neutron number with maxima at N=151. As a 
result of following configurations with a fixed  K$_{0}$ quantum 
number, the ground state fission paths obtained for the odd-mass nuclei 
are always higher than the AV ones. This specialization energy effect 
together with the pronounced quenching of neutron pairing correlations 
(leading to an enhancement of the  GCM and/or ATD collective masses as 
compared to the AV ones) are fully taken into account via the 
Ritz-variational minimization of the HFB-EFA energies. 

The present calculations represent a first step towards a better
understanding of the physics of spontaneous fission in odd-mass nuclei 
and there is still room for further improvements. For example, the evaluation of the 
collective inertias relies  on approximations that are not fully 
understood in the finite temperature case which is used to justify the 
formulas within our EFA methodology. This lack of full understanding 
leads to phenomenological prescriptions like the one used here to avoid 
potential divergences when two-quasiparticle energies cross. Another 
relevant aspect is the kind of treatment used to find the fission path: 
a dynamical treatment favors stronger pairing correlations than the 
least energy one employed in the present study
and this could substantially change not only the 
specialization energy but also the collective inertias. Finally, the lowering
of the first barrier due to triaxiality and its impact on the spontaneous
fission life-time in the case of odd-nuclei has still to be elucidated. There is no 
doubt that all these, still unresolved, aspects of the spontaneous 
fission in odd-mass nuclei deserve further studies in the future. Work along these lines 
is in progress and will be reported elsewhere.

\begin{acknowledgments}
The  work of LMR was supported by the 
Spanish grants FIS2012-34479-P MINECO, FPA2015-65929-P MINECO and FIS2015-63770-P MINECO.
\end{acknowledgments}

\end{document}